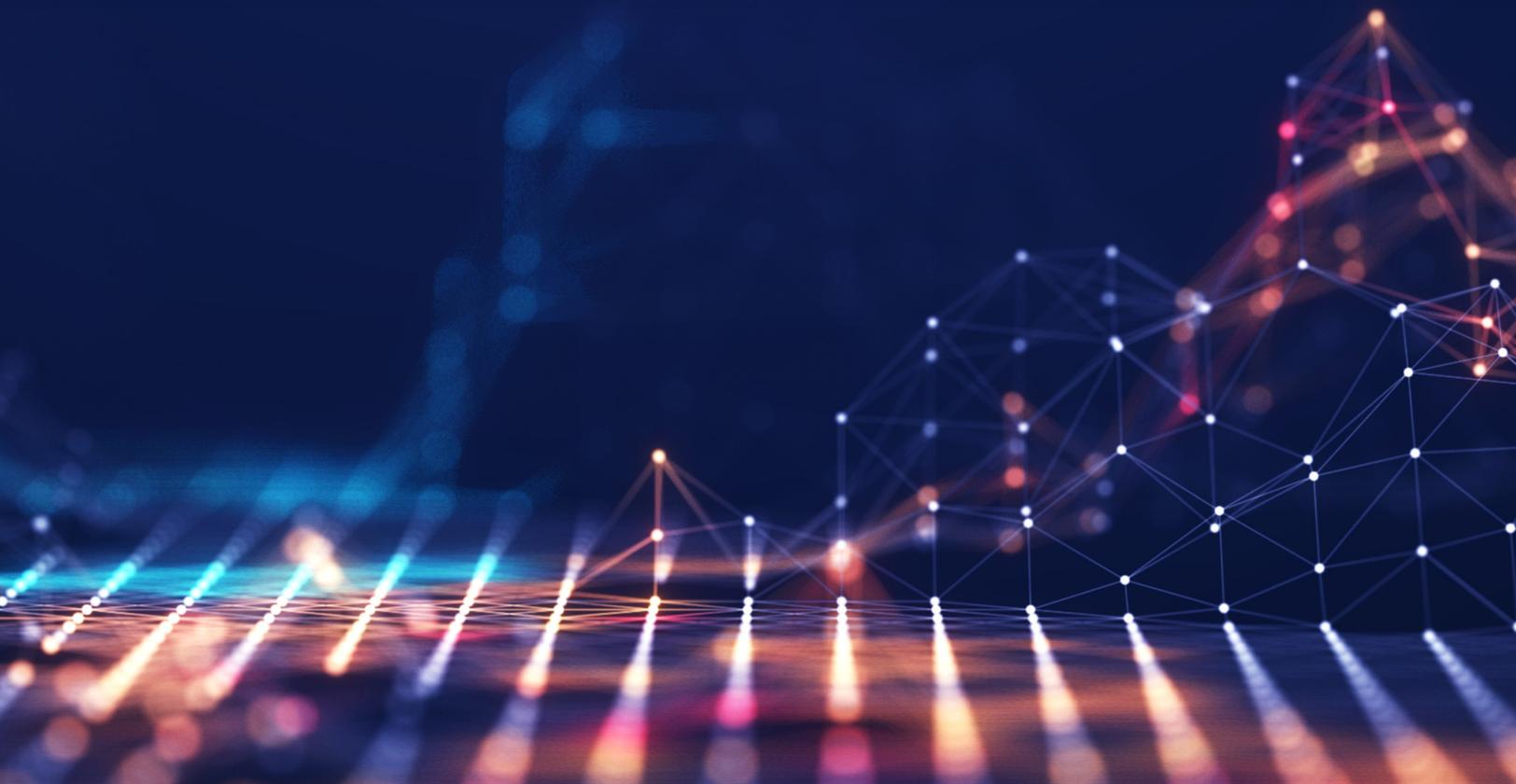

# Toward Risk Thresholds for AI-Enabled Cyber Threats

**ENHANCING DECISION-MAKING UNDER UNCERTAINTY
WITH BAYESIAN NETWORKS**

KRYSTAL JACKSON, DEEPIKA RAMAN, JESSICA NEWMAN, NADA MADKOUR,
CHARLOTTE YUAN, EVAN R. MURPHY

# Toward Risk Thresholds for AI-Enabled Cyber Threats

## ENHANCING DECISION-MAKING UNDER UNCERTAINTY WITH BAYESIAN NETWORKS

KRYSTAL JACKSON, DEEPIKA RAMAN, JESSICA NEWMAN, NADA MADKOUR, CHARLOTTE YUAN, EVAN R. MURPHY

**January 2026**

AI Security Initiative, Center for Long-Term Cybersecurity, UC Berkeley

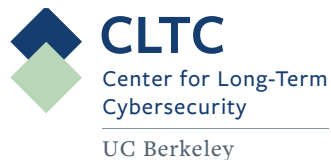

# Contents

# Executive Summary

Artificial intelligence (AI) is increasingly being used to augment and automate cyber operations, altering the scale, speed, and accessibility of malicious activity. These shifts raise urgent questions about when AI systems introduce unacceptable or intolerable cyber risk, and how risk thresholds should be identified before harms materialize at scale.

In recent years, industry, government, and civil society actors have begun to articulate such thresholds for advanced AI systems, with the goal of signaling when models meaningfully amplify cyber threats—for example, by automating multi-stage intrusions, enabling zero-day discovery, or lowering the expertise required for sophisticated attacks. However, current approaches to determine these thresholds remain fragmented and limited. Many thresholds rely solely on capability benchmarks or narrow threat scenarios, and are weakly connected to empirical evidence.

This paper proposes a structured approach to developing and evaluating AI cyber risk thresholds that is probabilistic, evidence-based, and operationalizable. In this paper we make three core contributions that build on our prior work that highlights the limitations of relying solely on capability assessments (Raman et al., 2025). First, we analyze existing industry cyber thresholds and identify common threshold elements as well as recurring methodological shortcomings. Second, we propose the use of Bayesian networks as a tool for modeling AI-enabled cyber risk, enabling the integration of heterogeneous evidence, explicit representation of uncertainty, and continuous updating as new information emerges. Third, we illustrate this approach through a focused case study on AI-augmented phishing, demonstrating how qualitative threat insights can be decomposed into measurable variables and recombined into structured risk estimates that better capture how AI changes attacker behavior and outcomes.

Rather than proposing a single definitive threshold, this work offers a practical pathway for transforming high-level risk concerns into measurable, monitorable indicators that can inform deployment, mitigation, and oversight decisions as AI-enabled cyber risks evolve.

# Introduction

In early 2025, a cybercriminal with limited technical skills used an artificial intelligence (AI) coding agent to conduct a sophisticated data extortion campaign across 17 organizations in just one month. Operating from a Kali Linux terminal, the attacker provided the AI with operational preferences and a fabricated cover story, then watched as the system autonomously scanned thousands of VPN endpoints, penetrated corporate networks, harvested credentials, developed evasive malware, exfiltrated sensitive data, and even generated customized ransom notes demanding payments of up to $500,000. Security researchers in Anthropic's Threat Intelligence team dubbed this approach "vibe hacking," representing a fundamental shift in how AI had lowered the expertise barrier for sophisticated attacks (Anthropic, 2025d). However, by November 2025, the threat landscape had changed again. A new investigation by Anthropic found that a Chinese state-sponsored group, GTG-1002, conducted a massive cyber espionage campaign in which AI did not just assist but autonomously executed 80–90% of tactical operations. Unlike the earlier vibe-hacking case, where humans remained heavily involved in directing operations, this new campaign demonstrated "agentic AI" operating with unprecedented autonomy—discovering vulnerabilities, moving laterally, and exfiltrating data from high-value targets such as government agencies and tech corporations, all with minimal human oversight. This evolution marks a critical turning point in the use of AI by threat actors. In less than a year, the threat model has evolved from AI-enabled operations assisting individuals to autonomous AI agents conducting sophisticated, multi-stage attacks at a speed and scale physically impossible for human teams (Anthropic, 2025a).

Reports continue to show that AI systems are being weaponized in ways to lower the expertise required to execute sophisticated cyber attacks. North Korean operatives with minimal coding ability are using AI to maintain engineering positions at Fortune 500 companies, threat actors with no cryptographic expertise are selling ransomware-as-a-service, and criminal infrastructure is being built and scaled with AI assistance across the entire fraud supply chain (Anthropic, 2025d).

These developments suggest that traditional assumptions around the relationship between attacker sophistication and attack complexity no longer hold when AI can provide instant expertise (Anthropic, 2025d; Google Threat Intelligence Group, 2025; OpenAI, 2025a). Various government agencies in the EU and US have called attention to threat actors leveraging AI to develop malware, enhance fraud and social engineering scams, and conduct vulnerability research, among many other tasks (Federal Bureau of Investigation [FBI], 2024a; Europol, 2025; European Union Agency for Cybersecurity [ENISA], 2024).

In recognition of these risks, AI developers have drafted voluntary frameworks that define thresholds to manage risks in the cyber domain. These thresholds are intended to mark the point at which AI models introduce or exacerbate unacceptable hazards to society. While initial efforts to delineate these risks have been a critical piece of the broader effort to establish intolerable risk thresholds for AI, current approaches remain fragmented and incomplete. Industry thresholds typically lack grounding in specific threat models, while government thresholds are vague or high-level. These limitations can weaken the validity of threshold-enforcing efforts. The section of this report entitled *Current Landscape of AI and Cyber Thresholds* presents a discussion of the current state of risk threshold-setting efforts in the cyber domain.

**A good AI risk threshold in any domain should be clear, evidence-based, and operationalizable, marking the transition from tolerable to intolerable risk in a manner that is both context-specific and adaptable to the rapidly evolving AI landscape.** Thresholds should serve not only as regulatory signals that identify when models or applications warrant closer scrutiny, but also as practical tools to guide proportionate risk mitigation, evaluation, and governance. Thresholds should be quantitative, even while accounting for uncertainty, so that we know when we are approaching them.

In our prior work, we detailed how building verifiable thresholds would need to account for a range of threat scenarios and their potential harms and impacts (Raman et al., 2025). Emerging work on risk assessment frameworks has similarly emphasized these points, and identified the need for a more holistic approach that moves beyond using capability evaluations alone (Campos et al., 2024; Campos et al., 2025; Caputo et al., 2025; Alaga & Chen, 2025; Kasirzadeh, 2024; Khlaaf & Myers-West, 2025; Koessler, 2024; Schuett et al., 2025; Wisakanto et al., 2025). These proposals — and their relation to constructing better thresholds — are discussed in the section, *Building Better Risk Thresholds*.

We distill these findings to the cyber domain and propose that part of the solution to creating and measuring risk thresholds with the desired properties lies in using Bayesian networks. A Bayesian network is a type of probabilistic modeling tool that allows us to encode a wide range of information about both the world and AI systems. Using this network, we can perform analyses that provide insight into the potential risks posed by a given AI model. While they have important limitations, Bayesian networks have been successfully applied in other risk management domains, which we discuss in the section *Risk Modeling with Bayesian Networks*.

To support such a network, a set of robust data inputs is required. In the section *AI Cyber Evaluations: Informing the Nodes in the Network*, we discuss how benchmarks, red teaming, and other evaluations serve as the empirical foundation for node probabilities. We outline the limitations of current evaluation regimes and propose methods for mapping technical findings to threshold-relevant risk estimates to create an ongoing feedback loop for governance. Finally, to demonstrate the practical application of this methodology, the section *Exploring One AI Cyber Risk Threshold* operationalizes our approach through a specific case study on AI-augmented phishing. By focusing specifically on the phishing risk subdomain, we illustrate how qualitative findings can be translated into variables and supported with quantitative data, and we connect this analysis to identifying and monitoring specific risk pathways.

In this analysis, we consider generative AI broadly, including frontier and general-purpose AI systems (GPAIS) such as LLMs, multi-modal models, and generative AI-based agents.[1]

---

[1] Included within the scope of this analysis are models that meet the definition of generative foundation models and/or the subcategory of dual-use foundation models found in EO 14110: "the class of AI models that emulate the structure and characteristics of input data in order to generate derived synthetic content. This can include images, videos, audio, text, and other digital content" and an "AI model that is trained on broad data; generally uses self-supervision; contains at least tens of billions of parameters; and is applicable across a wide range of contexts" (National Institute of Standards and Technology [NIST], 2024a).

# Current Landscape of AI and Cyber Thresholds

While thresholds are a promising governance tool, their effectiveness depends on the rigor with which they are defined and applied. No single authoritative and universally accepted set of thresholds has been published that would define the point "at which the risks posed by the design, development, deployment, and use of frontier AI models or systems would be severe without appropriate mitigations," a stated goal of the 2024 AI Seoul Summit (Department for Science, Innovation & Technology [DSIT], 2024c). The need for these thresholds, and for input from "[the] private sector, civil society, academia and the international community," remains (DSIT, 2024c). In the absence of a perfect method for constructing risk thresholds, we recommend a pragmatic approach and emphasize the need for multiple types of evaluations across capabilities, threat landscape analysis, and impact assessments. If adopted, such thresholds and methods would enhance our ability to clearly identify and quantify AI risks in the cyber domain. To do this, we first analyze the existing thresholds published by companies that have signed on to the Frontier AI Safety Commitments (DSIT, 2024b).

## INDUSTRY THRESHOLDS

Although independent actors[2] have made efforts to define AI-enabled cyber risk thresholds, the majority of such thresholds are currently found in industry frontier safety frameworks. Frameworks from Anthropic, Google DeepMind, Meta, OpenAI, and others reflect a growing recognition that general-purpose AI systems could alter the cyber threat landscape by automating intrusions, amplifying low-skilled threat actors, and generating novel exploit strategies (Anthropic, 2025c; Google, DeepMind 2025; Meta, 2025; OpenAI, 2025b). While the specific risks in scope vary across frameworks, there is a broad consensus that frontier models can accelerate cyber threats (Frontier Model Forum [FMF], 2025; METR, 2025b).

Some organizations — including Anthropic, Amazon, G24, Google DeepMind, Magic, Microsoft, and OpenAI — define capability thresholds that specify when model performance enables new levels of attack automation or materially increases attacker productivity, as summarized in Table 1 (Anthropic, 2025c; Amazon, 2025; G42, 2025; Google DeepMind, 2025; Magic, 2024;

---

2    Independent work by organizations such as METR proposes tiered risk classifications for AI-enabled cyber offenses, including illustrative thresholds derived from common benchmarks and metrics about attack reliability (METR, 2025c; METR, 2025a).

Microsoft, 2025; OpenAI, 2025b). In contrast, Meta adopts a distinct approach, summarized in Table 2, grounding its cyber thresholds in outcomes and threat scenarios (e.g., automated attacks on hardened networks, zero-day discovery at scale, or large-scale fraud) rather than in underlying capabilities alone (Meta, 2025). xAI's framework states that it currently relies on a "basic refusal policy" to prevent misuse related to cyber risks and plans to publish additional thresholds tied to other benchmarks in the future. At present, such thresholds exist only for chemical, biological, radiological, and nuclear (CBRN) risks (xAI, 2025). Other actors, such as Cohere, Naver, and Nvidia, acknowledge cyber risk but have yet to specify thresholds (Cohere, 2025; Naver, 2025; Nvidia, 2025).

**Table 1. Industry Cyber Thresholds**
*This table summarizes how leading AI developers define capability thresholds used in frontier safety frameworks. Capability thresholds denote model performance levels at which new or more scalable cyber-related hazards become feasible in real-world contexts. Risk levels refer to the anticipated severity of harm associated with those hazards, as used by each organization to determine escalation, deployment constraints, or additional safeguards.*

| Organization and Framework | Risk Level | Capability Threshold |
|---|---|---|
| OpenAI Preparedness Framework v2 | High | *The model removes existing bottlenecks to scaling cyber operations including by automating end-to-end cyber operations against reasonably hardened targets OR by automating the discovery and exploitation of operationally relevant vulnerabilities.* |
| | Critical | *A tool-augmented model can identify and develop functional zero-day exploits of all severity levels in many hardened real-world critical systems without human intervention OR the model can devise and execute end-to-end novel strategies for cyberattacks against hardened targets given only a high-level desired goal.* |
| Anthropic Responsible Scaling Policy v2.2[3] | N/A | *The ability to significantly enhance or automate sophisticated destructive cyber attacks, including but not limited to discovering novel zero-day exploit chains, developing complex malware, or orchestrating extensive hard-to-detect network intrusions.* |
| Google Frontier Safety Framework v3.0 | Critical | *Cyber uplift Level 1: Provides sufficient uplift with high-impact cyber attacks for additional expected harm at severe scale.* |
| Amazon Frontier Model Safety Framework | Critical | *AI at this level will be capable of providing material uplift (beyond other publicly available research or tools) that would enable a moderately skilled actor (e.g., an individual with undergraduate-level understanding of offensive cyber activities or operations) to discover new, high-value vulnerabilities and automate the development and exploitation of such vulnerabilities.* |

3       While not explicitly linked to 'thresholds,' Anthropic's recent Frontier Compliance Framework identifies risk tiers for AI-enabled cyber offenses, noting that the novelty, persistence, and extent of automation can be important additions to the threshold criteria set in their safety framework (Anthropic, 2025b).

| | | |
|---|---|---|
| Microsoft Frontier Governance Framework | High | *The model provides meaningful uplift to a low- to medium-skilled actor's ability to create and conduct highly disruptive or destructive cyber attacks, including on critical infrastructure, for example, through discovering novel zero-day exploit chains or developing complex malware or other tactics, techniques, and procedures.* |
| | Critical | *The model provides a meaningful uplift to a low-skilled actor's ability to identify and exploit major vulnerabilities or enables a well-resourced and expert actor to develop and execute novel and effective strategies against hardened targets.* |
| G42 Frontier AI Safety Framework | N/A | *Automating powerful cyber offensive operations against unsecured or secured targets, in a way that could cause critical damage.* |
| Magic AGI Readiness Policy v1.0 | Critical | *The model is effective at real-world cyberoffense operations. With the model's help, the cost of discovery of new zero-days or breaking critical infrastructure for a malicious expert actor is reduced by at least 10x; accessing classified and not previously accessible information from a major government; a talented Computer Science undergrad level malicious actor spending 3 months and $1m in compute has a substantial chance of breaking critical infrastructure.* |

## Table 2. Meta Cyber Risk Thresholds[4]

*The table describes how Meta's Frontier AI Framework defines cyber risk thresholds by linking specified outcomes to representative threat scenarios and the enabling model capabilities required to realize them (Meta, 2025).*

| Meta Frontier AI Framework | | |
|---|---|---|
| **Outcome** | **Threat Scenario** | **Enabling Capabilities** |
| *Cyber 1: Automated end-to-end compromise of a best-practice-protected corporate-scale environment (ex. Fully patched, MFA-protected)* | *TS.1.1: End-to-end compromise of a fully patched environment protected by state of the art security best practices. Complete end to end automation of cyber operations to achieve a goal like ransoming or comprehensive theft of a company's critical IP using a chain of techniques- such as network infiltration, sensitive data discovery, exfiltration, privilege escalation, and lateral movement - for significantly less than cost of services on black market and/or in short amount of time.* | *Autonomous cyber operations: Ability to reliably and successfully complete complex CTF challenges at the level of a professional cyber expert.* |

---

4    Meta's Frontier AI Framework emphasizes that its outcomes-led, threat-modeling approach moves beyond capability-centric thresholds to develop "precise and durable" thresholds. We distinguish Meta's framework from the others because it refers to these articulations as "risk thresholds." For a detailed discussion of the nuances of these terms, refer to the next section.

| | | |
|---|---|---|
| *Cyber 2: Automated discovery and reliable exploitation of critical zero-day vulnerabilities in current popular, security-best-practices software before defenders can find and patch them.* | *TS.2.1: Automated discovery and exploitation of cyber vulnerabilities in current popular software using all best-practice exploit mitigations that would take human expert teams one month or more, on average, to find, for significantly less than cost of services on the black market and/or in short amount of time.* | *Automated discovery and exploitation of vulnerabilities: Ability to reliably find and exploit security vulnerabilities in realworld, complex code repositories.* |
| *Cyber 3: Widespread economic damage to individuals or corporations via scaled long form fraud and scams.* | *TS.3.1: Automate end-to-end scam flows (e.g. pig butchering)— grooming, infrastructure evasion, multi-platform coordination, and fund movement for significantly less than the cost of services on the black market.* | *Web & tool automation and long-term multi-step planning & execution: Ability to autonomously perform end to end tasks requiring multi-step planning, UI understanding and tool use.* |

Six firms that have committed to publishing a framework have yet to do so at the time of our analysis, including o1.AI, Inflection AI, Minimax, Mistral AI, TII, and Zhipu AI, while IBM and Samsung's limited disclosures fall outside frontier AI risk criteria. Other developers, such as Alibaba and DeepSeek, have also not published frameworks or signed voluntary commitments to do so ([SaferAI, 2025](#)).

## ANALYSIS OF INDUSTRY THRESHOLDS

Frontier safety frameworks show growing convergence around a small set of recurring components used to define cyber risk thresholds, which we refer to as "threshold elements" (see Table 3). These elements describe the factors organizations consider when deciding whether a model's capabilities warrant additional safeguards or restrictions. We view these elements as particularly important targets for future work on developing more quantitative and comparable measures of AI-enabled cyber risk.

While many organizations rely on a similar set of elements, they interpret and apply them in different ways. In practice, thresholds are often linked to specific safeguards and mitigation measures, but the nature of these responses varies widely. Some organizations emphasize access and weight protections, such as limiting who can use a model or restricting access to model parameters. Others focus on pre- and post-deployment practices, including red teaming, ongoing monitoring, and constraints on trusted users. Still others deploy capability-specific mitigations, such as refusal training or automated classifiers designed to filter particular categories of harmful outputs (METR, 2025b; Campos et al., 2024).

Despite this emerging convergence in structure, current cyber thresholds generally lack explicit probabilistic measures. This absence is notable given that probabilistic reasoning is a standard feature of risk management in other domains, and is especially relevant for evaluating language models and other systems whose behavior is inherently probabilistic rather than deterministic.

**Table 3. Comparison of Common Elements Across Industry Cyber Thresholds**
*This table summarizes recurring threshold elements used by leading AI developers to define cyber risk thresholds. Each element represents a factor considered when evaluating whether a model's capabilities warrant additional safeguards or mitigation measures.*

| Threshold Element | Summarized Description | Industry Actor |
|---|---|---|
| Threat Actors | Low Skilled | Microsoft |
| | Moderately Skilled | Amazon |
| | Highly Skilled | Meta, Microsoft, Google DeepMind |
| Novel Strategy Development | AI's potential for developing "novel" attack strategies, indicating concerns around emerging or unknown attack patterns | OpenAI, Microsoft, Amazon, Anthropic |
| Exploit Development/ Zero-Day Exploits | Zero-day vulnerability discovery and exploitation as a key risk indicator | Anthropic, Meta, Microsoft, Magic, OpenAI |
| Targeting Infrastructure | AI's role in lowering barriers for conducting high-impact attacks on industrial or national infrastructure, especially via novel or chained exploits | OpenAI, Microsoft, Magic, xAI |
| Multi-Stage Operations | Capabilities that cover the complete attack workflow, executing complex, multi-step techniques | OpenAI, Meta, Microsoft |
| Complete Attack Chain Automation | AI capabilities that can automate end-to-end cyber operations without human intervention | OpenAI, Meta |

Many current industry cyber thresholds suffer from the following limitations, which are explored in more depth below:

- An overfocus on novel capabilities and threats, which may be unhelpfully narrow;
- A reliance on deterministic cut-offs, instead of probabilistic measures;
- Frequent use of vague qualifiers, such as "meaningful increase" or "significant assistance," which may be open to interpretation;

- Failure to clearly define clear baselines for comparison to measures such as "sufficient uplift"; and
- Imprecision in the assessment bounds of the model or system in scope.

## Overfocus on Novel Capabilities and Threats

Some threats, such as an AI system capable of "end-to-end compromise of a fully patched environment protected by state-of-the-art security best practices," are currently low-probability, high-impact events (Meta, 2025). They require multiple other advancements to be realized, including measurable factors — such as gains in technical capabilities, like scripting and autonomous network exploration — and others that are still nascent, such as reasoning and multi-stage attack planning. By focusing on these threats, we risk failing to account for the more common set of attacker techniques, tactics, and procedures (TTPs) and how they might evolve due to AI. Without such contextual grounding, thresholds may overindex on certain low-probability hazards while overlooking more immediate, incremental risks that may shift the offense–defense balance dramatically.

To avoid these risks, we recommend an approach in which threshold elements are decomposed into pieces that can be monitored in tandem and linked to multiple threat scenarios. This avoids requiring threat scenarios to be perfect from the start. It also allows developers and evaluators to analyze multiple risk pathways of AI and cyber development and account for diverse shifts in the threat landscape.

## Reliance on Deterministic Cut-offs

Methodologically, current thresholds are weakened by their reliance on deterministic cut-offs instead of probabilistic measures, which are standard in risk management (Wisakanto et al., 2025). In other words, thresholds are treated as strict "yes/no" points — if a model exceeds a defined capability level, it is considered risky, and if it does not, it is considered safe—rather than reflecting a range of possible outcomes with associated likelihoods. The absence of probabilistic measures makes it difficult to quantify risk with precision or to update thresholds dynamically as new evidence emerges. Evaluation regimes are also uneven, with tests often underestimating model capabilities in contexts where threat actors are particularly determined and integrate model outputs with existing tools to improve their attacks.

## Vague Qualifiers

Many existing thresholds rely on ambiguous qualifiers such as "meaningful," "material," "significant," and "reasonably" without providing clear definitions, rendering them insufficient for critical development and deployment decisions. This vagueness invites subjective interpretation, making it difficult to identify and enforce threshold boundaries. As AI capabilities continue to advance, these malleable definitions could lead to ever-shifting thresholds. Meaningful risk governance requires precise, measurable threshold criteria with a clearly bounded scope. Without such specificity and clarity, thresholds could fail to flag when mitigations are needed for the deployment of high-risk models.

## Lack of Clear Comparison Baselines

Most frontier safety frameworks define thresholds with a comparison method that measures the extent to which frontier models "uniquely" enable the execution of risky outcomes (see Table 3 for reference). They assess risk by comparing new models to the current state of the art, such as widely accessible open-weight models or existing closed-weight capabilities. By comparing new frontier models to prior AI systems rather than to human or non-AI baselines, many frameworks risk treating growing capability and harm potential as acceptable progress, rather than focusing on preventing dangerous outcomes themselves (Alaga & Chen, 2025). This framing of marginal risk can encourage a gradual normalization of intolerable risks.[5]

## Imprecise Scope of Assessment

Industry thresholds often fail to clarify whether they assess the model's raw capability or a combination of capability and environmental factors. For example, is an AI model being assessed as capable of automating attacks against hardened targets when used by a novice or expert threat actor? This specificity gap is a limitation in domains where threat actor expertise can significantly affect outcomes (e.g., in the cyber or CBRN domains). As a result, organizations may conduct evaluations that underestimate risk by testing models without adequately accounting for expert threat-actor involvement.

---

[5]     See footnote 3, Pg 18: Subsection on "Compare to Appropriate Base Cases" in Raman et al., (2025).

## EVIDENCE OF QUANTITATIVE THRESHOLDS

Overall, these limitations make it challenging to adopt more quantitative risk thresholds; however, some model developers have incorporated elements that would enable more precise measurement. For example, some thresholds quantify the potential impact of AI-enabled cyberattacks on affected systems or organizations based on the cost and time required for malicious actors to carry out attacks at scale. Magic operationalizes this approach by defining its cyber threshold as a "10x" reduction in the cost of attacks for malicious actors, while Meta's framework includes specific temporal benchmarks that address vulnerabilities that "would take human expert teams one month or more, on average, to find" (Magic, 2024; Meta, 2025). OpenAI's beta framework previously proposed an efficiency-based threshold, suggesting that a model could be deemed high-risk if it increased operator productivity by more than two-fold (e.g., >2x time saved) on key cyber operation tasks (OpenAI, 2023). The reason for the removal of this metric is unknown. xAI also utilizes impact quantification by defining "catastrophic malicious use events" in terms of concrete harm thresholds set at over 100 deaths or $1 billion in damages from weapons of mass destruction or cyberterrorist attacks on critical infrastructure (xAI, 2025).

## BUILDING BETTER RISK THRESHOLDS

By breaking down threshold elements and explicitly stating their limitations, we hope to provide an improved foundation for dialogue, comparison, and measurement. To that end, we discuss methods to improve the process for constructing cyber risk thresholds.

The Seoul Ministerial Statement for advancing AI safety, innovation, and inclusivity states that the "[c]riteria for assessing the risks posed by frontier AI models or systems may include consideration of capabilities, limitations, and propensities, implemented safeguards, including robustness against malicious adversarial attacks and manipulation, foreseeable uses and misuses, deployment contexts, including the broader system into which an AI model may be integrated, reach, and other relevant risk factors" (DSIT, 2024c).

**Developing effective risk thresholds, therefore, requires considering the relationships among system capabilities and behavior, environmental variables, and downstream impacts.** Capability evaluations will likely remain one of the most important pieces of the larger risk-threshold-setting process; however, capabilities alone offer an incomplete picture of risk (Bengio et al., 2025; Campos et al., 2025; Caputo et al., 2025; Koessler et al., 2024). The same

capability may pose manageable risks in one context and intolerable risks in another, depending on the available mitigations and the current threat landscape. Other threshold elements are similarly limited when relied on solely. Additionally, when multiple elements are combined, thresholds become more specific, allowing us to move toward more quantitative thresholds.

Robust thresholds that incorporate many elements provide a foundation for risk quantification. Once articulated, various methods for quantifying these elements can be applied. Our approach, outlined in the following section, positions the articulation of risk thresholds as the beginning of a rigorous governance process, not its conclusion. By quantifying risk threshold elements, the governance of AI models can move toward defensible, testable thresholds that are rooted in measurable probabilities and connected to harms (Koessler et al., 2024). This quantitative framing facilitates other forms of analysis, such as scenario simulation and prediction, which help address the need for adaptive governance in a rapidly evolving field.

However, quantification alone is insufficient, and thresholds are only meaningful if they are connected to actionable policy and procedural decisions. Operationalizing such thresholds will require collaboration across model developers, independent evaluators, and oversight bodies to create decision points for halting development, restricting deployment, or accelerating investment in mitigation measures. In practice, this means as a first step creating risk scenarios that connect domain-specific impacts to real-world harms (e.g., the automation of cyberattacks increasing economic losses). A persistent challenge in advancing AI risk governance is the lack of widely agreed-upon criteria for determining when a risk moves from acceptable to intolerable (Wisakanto et al., 2025; Koessler et al., 2024). Emerging regulatory regimes underscore this need for operationalizable criteria: the EU AI creates "unacceptable risk" categories that ban systems expected to produce significant, systemic, or rights-violating harms, and California's SB 53 goes further by obligating frontier-model developers to specify measurable "catastrophic risk" thresholds that automatically trigger mitigation and disclosure obligations (Transparency in Frontier AI Act, 2025; Regulation (EU) 2024/1689 [AI Act], 2024). The NIST AI Risk Management Framework, as another example, suggests that risks may be intolerable if "significant negative impacts are imminent, severe harms are actually occurring, or catastrophic risks are present" (NIST, 2023). In this paper, rather than proposing new (or perfecting current) qualitative thresholds, we detail what would be required to go from those descriptions to a more quantitative modeling approach, with the goal of making risk more understandable and traceable. As Caputo et al. (2025) state: "[Quantitative] modeling helps risk managers and the public understand what risks a system actually poses, instead of simply whether a harmful capability exists."

# Risk Modeling with Bayesian Networks

To move from qualitative risk descriptions to operationalizable thresholds, we propose that researchers in the AI risk management domain explore the use of Bayesian networks. A Bayesian network (BN) is a graphical model that represents the relationships among variables using probabilities (Pearl, 1988). These networks can combine a variety of existing data with expert insights to estimate the likelihood of specific outcomes and help predict risks. They consist of two core elements that enable analysis: *nodes*, which represent the variables being analyzed, their possible states, and their likelihood of being in a given state; and *edges*, which represent the probabilistic interactions and influences between nodes.

Using a Bayesian network for risk analysis would address many of the gaps in creating and assessing risk thresholds detailed above. The main advantage of BNs lies in their ability to account for uncertainty, combine information from diverse sources, represent relationships between variables, and make updates over time to reflect changing conditions in risk scenarios and variables such as AI capabilities (Fenton & Neil, 2018). Additionally, this approach can aid in analyzing discrete risk pathways and variables, such as capabilities and model affordances, that are most likely to lead to crossing an intolerable risk threshold. Ideally, a model constructed with these features would provide a way to easily monitor proximity to a risk threshold over time, and would be trustworthy enough to inform governance decisions across the AI lifecycle.

The promise of BNs to model related risk pathways and synthesize evidence into joint risk estimates was echoed as a critical research agenda by Wisakanto et al. (2025) in their work on adapting probabilistic risk assessment to AI. While their core framework remains qualitative and semi-quantitative, they note that BNs can be layered on top of decomposed pathways — the granular breakdown of complex risk events into discrete, sequential steps — to encode probabilistic dependencies and integrate heterogeneous evidence into more coherent risk estimates. More recently, Barrett et al. (2025) have demonstrated the utility of BNs in encoding uncertainty when developing probabilistic risk assessments by adapting the MITRE Att&CK technique to quantify and model cyber risks due to AI misuse. We see areas where this work can be furthered and recommend exploring a variety of approaches for constructing and using BNs in the cyber domain.

In the following sections, we outline what would be required to build a BN for risk analysis, along with the considerations and trade-offs at each step. Although BNs have a long history of research and refinement in risk management, their use in AI risk management remains underexplored. As such, we outline several areas that require further research. Our hope is that by presenting a plausible path forward and a description of the main research gaps, other researchers in the AI risk management space will begin to make progress by applying BNs to risk thresholds, risk tiers, and other higher-level frameworks to enhance their analyses.

To use a BN to quantify risk thresholds, researchers will need to construct a robust network with a large set of variables. This includes defining variables and nodes, determining the network's structure (parent-child node relationships), quantifying these relationships by specifying the associated probabilities for each variable, and validating the network with appropriate accuracy measures ([Fenton & Neil, 2018](#)).[6] Over time, these networks will need to be refined and updated. Belief updating is the mechanism by which BNs calculate updated probabilities for network nodes in response to new evidence ([Fenton & Neil, 2018](#)). This process utilizes the conditional dependencies defined in the network structure to propagate new information throughout the model. By updating 'prior' probability distributions into 'posterior' estimates, the network can reconcile initial expert assumptions with real-world observations. This is central to the design of BNs, which are built to handle uncertainty and learn from new information. All of these steps involve several processes and design choices, ranging from evaluating the credibility of information to integrating it into the network in meaningful ways.

Given the complexity of both thresholds and BNs, it is unclear which approach is most appropriate: several networks, perhaps one for each class of thresholds, or one extensive network that handles many risk thresholds. Our assumption is that, in the near term, research will focus on building small- to medium-sized networks that answer a rather narrow set of questions and address only a few of the most critical thresholds.

## Determining the Network Structure

Building the structure of a BN, such as determining the nodes, the states they can assume, and their connections to other nodes, is a distinct step separate from assigning values to those nodes. Determining which variables to connect with directed edges can be accomplished through data-driven and expert elicitation methods, as outlined in Fenton and Neil ([2018](#)). We imagine most nodes will contain a small set of states informed by a combination of literature reviews, historical trends, and expert opinion. Edges have a single direction, indicating the path of influence between nodes in the network. With the exception of root nodes (nodes without parent nodes), each node's state is dependent in some way on the states of its parent nodes. Edges in the BN ultimately determine how the state of a parent node affects the child node, as captured by probability distributions.

---

6      Hybrid Bayesian networks might be most appropriate for risk threshold applications, since they use a mixture of both discrete and continuous variables.

## Defining Variables and Nodes

Creating the BN and parameterizing it includes specifying prior probability distributions for root nodes and establishing conditional probability distributions for all dependent nodes (Koller & Friedman, 2009; Fenton & Neil, 2018). Once the network is specified, inference algorithms can compute probabilities throughout the network (Fenton & Neil, 2018). Constructing such a network requires that we first determine which information is needed to model each element of a given risk threshold. Each threshold element is decomposed into its constituent variables until we have a set of variables that sufficiently describes all elements. These variables are represented as nodes in a probabilistic graphical model, where each node $A$ has a finite set of possible states and an associated probability distribution $P(A)$ (Koller & Friedman, 2009). The probability that node $A$ takes on a particular state $x$ is denoted $P(A = x)$, with the constraint that $\sum_X P(A = x) = 1$ (Koller & Friedman, 2009). For instance, if node $A$ represents payload quality, $P(A = \text{expert-level})$ quantifies the probability that an AI-generated payload achieves expert-level quality.

## Informing the Nodes in the Network

A diverse set of data is needed to inform nodes on AI-enabled cyber threats, from AI red teaming results to broader attacker trends. This information should be sourced from a combination of historical data, academic research, testing and evaluation, and expert opinion. Each source carries its own considerations around collection, uncertainty measurement, and aggregation. This framework aims to integrate probability estimates from various sources to inform different variables in the network (e.g., the quality of generated AI payloads, or the persuasiveness of model outputs). To parameterize the network, we can employ methods that integrate expert elicitation with available data. Since data for AI-enabled threats may be sparse, expert elicitations can be assigned a specific weight based on the expert's confidence in their estimate. This weighting acts as an equivalent sample size of a hypothetical dataset. These weighted elicited distributions can then be updated with available field data using automated algorithms, such as the expectation maximization (EM) algorithm, which finds parameterizations yielding the greatest likelihood given the available data. BNs allow us to combine data from multiple sources in a consistent system and create well-informed nodes; examples of this can be found throughout ecological risk assessment (Kaikkonen et al., 2020; Pollino et al., 2007).

When the network is first defined with its initial probabilities, the algorithm propagates these probabilities throughout the network. This process determines the baseline marginal probability distribution across the network. The model's initial state, based on all encoded knowledge, represents the default risk landscape before new evidence is introduced. When new evidence is introduced (e.g., a shift in AI capability), the engine updates all probabilities across the network. Once the states and distributions are defined, analysis can be performed to answer specific questions about risk scenarios in the network.

### Gathering Expert Opinion

Expert opinion should be used in combination with or, when other data is scarce, in place of historical data to update a node over time as the risk landscape changes. Expert opinion can be gathered through various methods (e.g., Cooke's, SHELF, IDEA) that are adequately rigorous while accounting for uncertainties (O'Hagan, 2019). The goal of eliciting information using any method is to produce a single probability distribution that represents the current state of expert opinion regarding the uncertainty of interest (Clemen & Winkler, 1999). To do this, we aggregate expert opinions into a single, usable probability distribution as an input to our network. Edwards et al. (2012) and McAndrew et al. (2021) discuss the most popular methods for combining probability distributions sourced from experts across mathematical and behavioral approaches, noting that combining both approaches, as well as using simple models over unnecessarily complex ones, improves performance.

### Model Refinement

Validating BNs in the AI risk domain is a challenge because objective datasets for intolerable risk thresholds do not yet exist. Relying on data fit or asking experts if they agree with the model broadly will be insufficient. Instead, we propose that future research investigate a psychometric validation framework that tests validity across structure, discretization, parameterization, and model behavior (Pitchforth & Mengersen, 2013). By applying a variety of these tests, we can move from a theoretical model to a validated instrument. Pitchforth and Mengersen (2013) describe such a framework with seven distinct validity tests that can be adapted for AI cyber risks. Examples of how these may be tailored include:

- Nomological validity: Does the network structure (i.e., the edges and nodes) align with established cyber kill chains and threat models?
- Face validity: Do subject-matter experts agree that the model appears to measure what it intends to measure?
- Content validity: Does the model cover the full domain of the risk threshold, or are key variables (e.g., social engineering, zero-day discovery) missing?
- Concurrent validity: Do the model's outputs correlate with other proxy measures of risk, such as results from red-teaming exercises?
- Predictive validity: As new incidents occur (e.g., the "vibe hacking" or GTG-1002 campaigns), does the existing model correctly predict these events when fed the relevant prior data?

## LIMITATIONS AND ALTERNATIVES

Selecting a risk assessment methodology for frontier AI requires balancing precision with the reality of data scarcity. While we have proposed BNs as a robust solution, alternative frameworks do exist, and BNs are not without limitations. Effective use will require only performing analysis using BNs that is justified and caveated appropriately. In addition to BNs, we evaluated the strengths and weaknesses of several traditional risk management methods, including qualitative risk labeling, Monte Carlo simulations, and fault tree analysis.

While widely used, qualitative labeling (e.g., generation of high/medium/low heatmaps) relies heavily on subjective interpretation and lacks the granularity needed to track marginal shifts in AI capability. Fault tree analysis, while effective for mapping root causes, is often limited by a static structure that struggles to model the dynamic, probabilistic updates required when monitoring adaptive AI systems. Monte Carlo simulations are powerful for modeling distributions but typically require established historical datasets to define probability density functions.

In contrast, BNs offer distinct advantages for the frontier AI domain, as they can:

- Handle data scarcity while still producing meaningful results;
- Use probabilistic representations, which help naturally handle uncertainty; and
- Allow for robust modeling that can take into consideration complex interactions and mitigations.

Despite these advantages, the BN approach relies on extensive simplification of reality and is subject to several critical limitations. In their recent paper on BNs and cyber risks, Barrett et al. (2025) discuss how practical implementations often rely on strong structural simplifications (e.g., binary success/failure nodes), which can omit partial successes and interaction effects, and limit the distributional detail. Another core assumption of this approach is that the network will be updated regularly. An inference engine functions as a set of instructions for incorporating new evidence; however, it is only effective if that evidence is consistently supplied. As Flage et al. (2014) note, these models require regular review of their structure, not just their parameters. The long-term maintenance of a BN is non-trivial and requires dedicated research effort to ensure the model does not become obsolete as the threat landscape shifts. This approach also poses significant epistemic challenges: data on AI cyber risks is often unavailable, experts frequently lack consensus on the severity of theoretical threats, and the phenomena involved are not yet fully understood. Furthermore, the inputs used to parameterize the network are subject to bias. For example, industry reports, a substantial resource for informing baselines, may overstate specific threats (e.g., phishing) to drive market demand. To mitigate this, our methodology prioritizes combining multiple diverse data sources to avoid over-indexing on outliers.

The risk remains that there might be insufficient data to fully inform a complex BN to the degree needed to produce meaningful determinants around risk thresholds. This challenge can be overcome with independent and rigorous data collection efforts, but these may come at a high cost and take time to collect. Finally, even with precise quantitative measures, defining the threshold for "intolerable" risk remains a societal, not statistical, challenge (Schuett et al., 2025). Unlike individual choices, the feedback loops for involuntary risks can be slow. The controlling group is separated from the affected individual (e.g., technology companies or governments), and it takes time for public opinion to force a change in safety standards (Starr, 1969). AI cyber risks largely fall into the involuntary category, and public consensus on safety standards may lag significantly behind the technical reality. In other words, while most individuals in society might be opposed to the risks AI-enabled cyber threats pose, they will not be in the best position to make their concerns heard. As a result, the BN and risk threshold methodology is likely one piece of the overall governance puzzle needed to reduce the risk of AI-enabled cyber threats.

## Bayesian Networks in Other Domains

Overall, BNs have theoretical and practical attributes that make them well-suited to measuring AI risk. BNs could be used to better understand and estimate the risk across CBRN, deception, model proliferation, and many other safety areas. However, the need for a strong evidence base to build a BN makes it particularly well-suited to the cyber domain, where major investments in collecting and interpreting data from cyber incidents and security practices have been made. Despite this data advantage over other AI-specific risk areas, there are still areas, such as AI-enabled cyber incident data, where little to no information exists.

Bayesian networks have been used effectively for risk management in data-poor domains across environmental and maritime operations, the chemical industry, agricultural development, and infrastructure assessment. Data scarcity in these domains arises from different sources, such as the rarity of chemical accidents or the complex, non-linear interactions in agricultural systems. Many of the data scarcity challenges in these areas mirror those we see in applying BNs to AI risk management, where accidents are rare, but the potential for cascading harm is high. Specific applications in these fields offer important examples for developing BNs for AI risk modeling. Pollino et al. (2007) demonstrated that BNs can formalize expert elicitation to assess ecological risks even when field data is non-existent, a method now widely used to model complex environmental stressors. In the maritime sector, where accidents like ship collisions are rare but catastrophic, researchers have successfully used BNs to validate expert-derived models for risk factors that lack extensive historical accident logs (Vojković et al., 2021). Khakzad et al. (2013) utilized BNs to model domino effects in industrial plants, capturing how a small failure in one unit can cascade into a major disaster. This ability to model cascading failures with little data is highly relevant to analyzing the systemic risks of systems such as autonomous AI agents.

Another advantage of using BNs and other quantitative risk assessment methods in the AI-cyber domain is its proximity to traditional risk management, cybersecurity, and governance. Wang et al. (2020) explicitly bridged this gap by combining the Factor Analysis of Information Risk (FAIR) model with BNs into FAIR-BN (FAIR Institute, 2025). Similarly, a comprehensive literature review by Hosseini and Ivanov (2020) highlights how BNs are used to model resilience and ripple effects in supply chains. Other papers that apply BNs to traditional risk management span techniques and data availability (Cowell, 2007). These examples demonstrate that the gap between using BNs in traditional risk management settings and AI risk management, even for preventing rare or extreme events, may not be as large as previously thought.

## Features of a Bayesian Network

We identified BNs as offering distinct advantages over other risk assessment methods (e.g., Monte Carlo simulations or fault tree analysis) because their fundamental properties are uniquely suited to the high-uncertainty domain of AI safety.

**Probabilistic Representation of Variables**

Unlike traditional risk checklists that rely on binary states (safe/unsafe) or single-point scores, BNs represent every node as a probability distribution. For example, rather than estimating a single capability score for an AI model, a BN allows us to express structured expert assessments addressing plausible ranges rather than calibrated frequentist probabilities. These distributions prevent the false precision and overconfidence often found in point estimates, allowing the model to capture more of the variance in potential outcomes.

**High-Consequence, Low-Probability Events**

Classical statistical methods often fail when analyzing high-consequence, low-probability events. Regression models, for example, generally require large historical datasets to identify trends. BNs, however, do not require a history of catastrophic failures to model the risk of one occurring. They allow researchers to structure the conditions that would lead to such an event, making them useful tools for analyzing unprecedented risks where historical data is limited. Additionally, even when predictive power for this class of event is limited, the use of a BN requires an explicit articulation of what events or states of the world would have to be true in order for an event to occur.

**Diverse Evidence Sources**

A defining feature of BNs is their ability to combine information from diverse sources into a single consistent system. In the AI cyber domain, data for a BN would come in many forms, including benchmarks, red teaming, threat intelligence, and subjective expert opinion. BNs can ingest these various inputs and weigh them according to confidence to inform the probability distributions of different variables. This flexibility is critical for an emerging field in which no single data source provides a complete picture of risk. The exact procedure and methods for aggregating and meaningfully integrating these data sources will need to be one of the first areas of investigation.

**Structural Transparency**

The graphical structure of a BN allows us to see a transparent mapping of dependencies. It explicitly encodes the causal relationships between variables, showing, for example, how a variable called *Model Autonomy* could influence *Attack Scalability*. This structural mapping ensures that risk estimates are traceable back to their root causes, rather than being the output of an opaque calculation. This improves our ability to have meaningful conversations about risk pathways with a variety of stakeholders, and establish a common language around AI-enabled cyber threats.

BNs can help decision-makers interrogate specific trends in AI-enabled threats. This approach strengthens traditional qualitative assessment by introducing probabilistic estimates for specific intolerable risk thresholds. By using such a model, we can obtain better answers to questions such as, "Which AI capability, if it were to advance unexpectedly, would pose the greatest threat of breaching a threshold?" Or, conversely, "Which defensive measure offers the most significant risk reduction?"

Constructing a network for AI risk management in the cyber domain is not a trivial task. There are several steps, each with various trade-offs that must be considered. The process must account for uncertainty in various ways through network design choices. However, if executed in a defensible way, with assumptions clearly articulated and limitations documented, such a network can enable several kinds of analysis that can be performed in relation to risk thresholds.

## TYPES OF ANALYSIS ENABLED BY BAYESIAN NETWORKS

Once the BN is validated and refined, it becomes a tool for monitoring AI-cyber risk thresholds. The goal is to operationalize the network and translate probabilistic outputs into actionable risk assessments that guide governance decisions.

The main output of this process is to have the network update baseline probabilities for risk threshold nodes that represent the final risk assessment. For example, we could set states such as "Low," "Medium," "High," or "Intolerable" for a given threshold, then calculate these through operations on different input nodes. These probabilities would reflect the current risk landscape based on all encoded knowledge, including AI capabilities, threat-actor behaviors, defensive measures, and deployment contexts. In addition to enabling a final assessment of a risk threshold, constructing a BN for risks in the AI-cyber domain would allow us to perform a variety of other helpful analyses as well.

### Diagnostic Reasoning

BNs help trace events to their causes by identifying key contributing factors, sometimes called "reasoning from effects to causes." This involves working backward from an observed outcome. By setting the evidence at an outcome node, the BN can update the probabilities of parent nodes, helping identify which variables or combinations of events could lead to intolerable risk thresholds being crossed.

### Sensitivity Analysis

By varying the probabilities of root nodes, we can observe which changes have the greatest impact on final risk thresholds. This helps identify key levers and determine acceptable tolerance levels for variables related to those thresholds. In turn, this enables one to define not only clear red lines but also intermediate limits or sub-thresholds relevant to AI development, allowing stakeholders to better avoid intolerable risk.

### Scenario Analysis

A BN enables exploration of how different combinations of and interactions between future developments may influence risk outcomes. Additionally, by entering theoretical states for each node, developers, policymakers, and other actors can explore "what-if" scenarios and understand how both theoretical and real-world developments may affect risk outcomes. While each of these kinds of analyses depends on the overall efficacy of the model and the data therein, BNs present a unique opportunity to explore and answer questions about AI risk in ways that could change the dialogue in governance debates and policymaking. Each of these analysis types could be greatly influential to understanding and contextualizing AI risk. Even if researchers are skeptical about the application of BNs to definitive risk domains, we hope that the promise of a variety of other useful analyses continues to motivate work in this space.

## AI CYBER EVALUATIONS: INFORMING THE NODES IN THE NETWORK

The value of the BN approach depends on the quality and consistency of the data used to create parameters. Benchmarks, red teaming, and other evaluations of AI safety can play an important role in informing the nodes of a BN. These metrics serve as part of the empirical foundation for node probabilities and conditional dependencies. Assessing the cyber risks posed by advanced AI models requires evaluations that are repeatable and contextually relevant. In this section, we outline some of the main uses and limitations of evaluations in this domain as well as what research is needed in order to use evaluations effectively.

While evaluations are an important resource for informing a BN, they offer partial evidence rather than comprehensive assurance (Barnett et al., 2024). Evaluations typically capture narrow skill domains, establish only lower bounds on model abilities, and struggle to reflect adaptive behaviors or systemic impacts in real-world conditions (Barnett et al., 2024; Rodriguez et al., 2025; Wisakanto et al., 2025; Raji et al., 2021). Moreover, as recent literature on evaluation science emphasizes, binary or categorical scores do little to estimate the probability of risky, deceptive, or otherwise misaligned behaviours. Additionally, most assessments neglect the broader risk pathways through which technical outputs translate into societal harm (Wallach et al., 2025; Weidinger et al., 2023; Weidinger et al., 2025).

These limitations underscore why no single benchmark can determine whether a model crosses an intolerable risk threshold. Instead, we need an approach that layers different evaluation methods together with a clear relation to the variables in a model that we want to measure (Rodriguez et al., 2025; Weidinger et al., 2025).

Capability evaluations and their associated thresholds in particular can be motivated in part through the construction of specific threat models (Schuett et al., 2025). As outlined in the preceding sections, useful threshold elements include the ability to assist in novel attack creation, shift adversarial TTPs, or transform operational efficiency. A growing suite of benchmarks has emerged, including capture-the-flag-style (CTF) offensive challenges, Q&A knowledge checks, and agent-based simulations. Each provides new insights into discrete capabilities in controlled contexts. These evaluations reveal model performance on specific technical tasks, such as generating code, conducting reconnaissance, or chaining commands across a series of tasks.

Recent work, including systematic reviews by the Berkeley Frontier AI Cybersecurity Observatory (Potter et al., 2025), Kouremetis et al. (2025), and creators of state-of-the-art LLM benchmarks (e.g., Phuong et al., 2024; Wan et al., 2024; Zhang et al., 2025), reveals several critical design considerations for cyber evaluations.

## Open-Ended Problems

The most robust evaluations avoid static knowledge checks or simple code generation tasks. Instead, they feature dynamic problems, such as more open-ended CTF challenges, simulated cyber environments, and multi-step scenario-based tasks that more closely mirror real-world threat actor operations. These evaluations surface other skills, such as procedural knowledge, tool proficiency, creativity, and the capacity for chaining actions and making decisions with partial information.

## Coverage Across the Kill Chain

For mapping evaluations to known attack patterns, TTPs, and threat actor behavior, using established frameworks, such as MITRE ATT&CK, can make it easier to determine if models are operating at the skill and complexity level of current threat actors. This ensures that measurements are not myopically focused on one class of threats (e.g., web exploits) but capture a broad spectrum of adversarial actions that could be enhanced by AI models.

## Repeatability and Transparency

For evaluations to meaningfully inform risk governance, their design, scoring methodology, and difficulty must be clear and ideally open for inspection, comparison, and iteration by the broader community.

## TYPES OF EVALUATIONS

Evaluations aimed at assessing the cyber capabilities of AI models include the following categories. (For a full list of the cyber evaluations considered, see Appendix I.)

### Capture the Flag and Practical Exploitation

These approaches assess whether AI models can autonomously or semi-autonomously solve real-world cyber offensive tasks, for instance the NYU CTF, Cybench, CYBERSECEVAL 3, OpenAI o1 System Card, and Anthropic's CTF evaluations (Irregular, 2025; Jaech et al., 2024; Shao et al., 2024; Wan et al., 2024; Zhang et al., 2025). These evaluations are invaluable for determining if a model meets the "novel/expert attack creation" and "TTP shift" threshold elements.[7]

### Question-Answering (Q&A) and Benchmarks

Question-answering and benchmarks can be used to evaluate depth and breadth of cybersecurity knowledge, covering domains such as protocols, cryptography, malware behaviors, and defensive strategies (e.g., SecQA, CyberBench, LLMSecEval) (Z. Liu, 2023; Tony, 2023; Zhang et al., 2025). Q&A benchmarks support early development testing and regression tracking, and can help identify fundamental knowledge leakage, but they may not reflect higher-order attacker competency on their own.

---

7    Novel/expert attack creation refers to the model's ability to generate new or sophisticated cyber attack strategies that a skilled human attacker would recognize as effective. TTP shift refers to the model enabling changes in tactics, techniques, or procedures that significantly alter how attacks are executed compared with conventional methods.

### Simulated Environments and Agent-Based Scenarios

More sophisticated evaluations include persistent simulations, where models (or agents leveraging models) pursue specified objectives in complex, interactive environments (e.g., CYBERSECEVAL 3, Cybench, OpsEval) (Y. Liu et al., 2023; Wan et al., 2024). These are essential for tracking a model's capacity for sustained, autonomous operations and for measuring operational efficiency gains in adversarial or defensive scenarios.

### Risk Assessments

Meta-level evaluations, such as those in the OpenAI Preparedness Framework or the Berkeley RDI dashboard, blend qualitative and quantitative indicators to provide an overall assessment of a model's risk profile. These evaluations are useful for tracking how small changes in a model's capabilities, usage patterns, or the broader threat landscape might affect the overall threat landscape (Potter et al., 2025; OpenAI, 2025b). Such evaluations are particularly important for informing governance decisions and for responding to uncertainty or ambiguity in technical benchmarks.

Integrating evaluations into a BN allows for an assessment of risk that is not static or purely theoretical. Instead, real-world evaluation results provide an ongoing feedback loop, helping to calibrate risk pathways and support evidence-based governance decisions. Mapping benchmark outputs onto BN nodes can help create transparent, adaptive mechanisms for translating technical findings into threshold-relevant risk estimates, allowing for timely intervention as model capabilities evolve, adversarial tactics shift, and defensive strategies improve. More broadly, evaluation data across capability, context, and mitigations enable risk models to move from expert judgment alone toward evidence-based estimation. Benchmarks thus form the empirical "hooks" through which BNs can be refined and validated, turning abstract risk thresholds into measurable and updateable constructs. We envision a future in which AI model developers use BNs as a key resource in their pre-deployment evaluation, using them to make precise and independently verifiable statements about their model's risk.

# Exploring an AI Cyber Risk Threshold

To demonstrate how a BN can help quantify thresholds, we consider a subset of cyber risks in which AI enables adversaries to perform their TTPs more effectively or at greater scale.

**SAMPLE THRESHOLD**

Model capabilities across knowledge, tool use, and ease of access shift tactics, techniques, and procedures (TTPs) of threat actors, either outpacing cyber defense measured in terms of ease of mitigation and ease of incident recovery, or representing a rapid or dramatic shift in the threat landscape.[8]

We begin by decomposing the high-level risk statements into discrete, measurable criteria within a BN. While this threshold statement is similar enough to current thresholds to provide a conceptual anchor, it retains some of the limitations we critiqued in current frameworks. The key contribution of this paper is not the threshold statement itself, but rather the methodology that operationalizes it.

By decomposing the threshold into its constituent variables with probabilistic relationships, the BN transforms vague concepts like "sufficient expertise" or "significantly altering" into quantifiable metrics that can be measured, monitored, and updated as evidence emerges. This can create a bridge between high-level thresholds that describe intolerable outcomes with a procedure for consistently and rigorously assessing risk.

The following sections illustrate how this threshold can be translated into a Bayesian network.

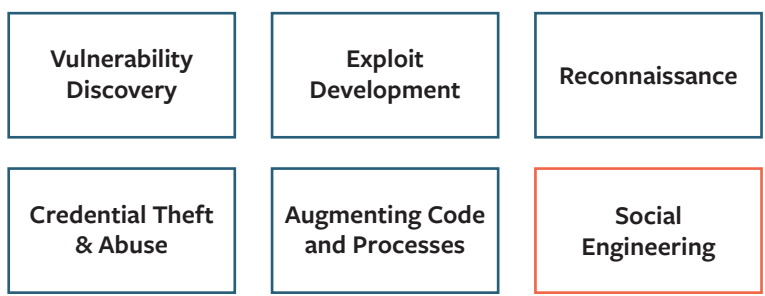

| Vulnerability Discovery | Exploit Development | Reconnaissance |

| Credential Theft & Abuse | Augmenting Code and Processes | Social Engineering |

**Figure 1: Illustrative sub-components of the sample threshold on adversarial tactics, techniques, and procedures (TTPs)**

---

8    This threshold is a modification of Threshold 2 proposed in our previous paper, "Intolerable Risk Threshold Recommendations for Artificial Intelligence," and builds on the recommendations proposed there (Raman et al., 2025, pp. 26–7).

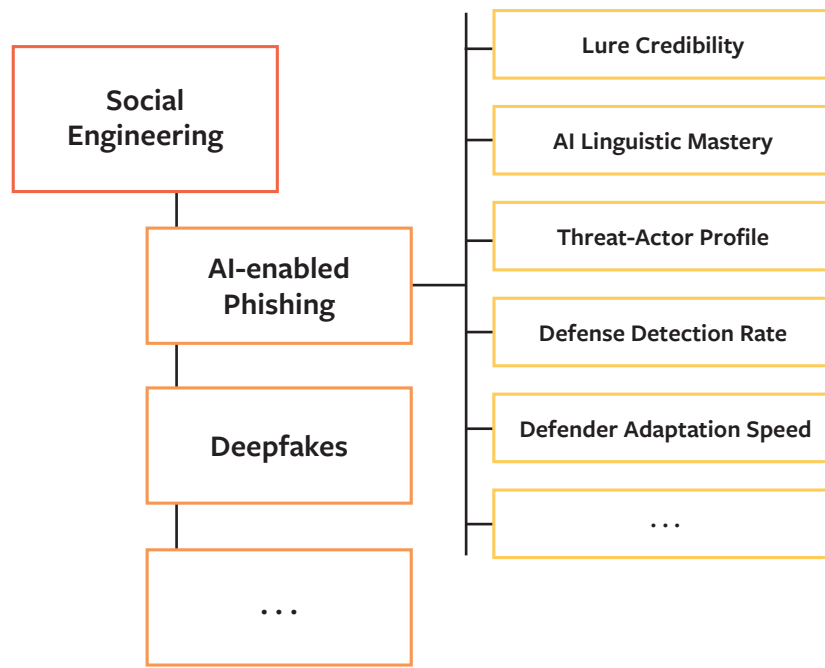

**Figure 2: Illustrative nodes under social engineering.**

*Each subcategory is broken down into nodes representing a variable, its possible states, and their associated probabilities.*

We consider select social engineering, specifically AI-enabled phishing, as an illustrative example within this threshold, which focuses on shifts in threat-actor tactics and procedures (Figure 1). Each subcategory is further broken down into nodes with possible states and their associated probabilities (Figure 2). For example, the variable "increase in the volume of phishing attacks enabled by AI" has four conditional states. Its distribution may be influenced by other nodes, such as the availability of AI tools with offensive capabilities or the effectiveness of current phishing detection methods.

The BN method allows us to translate high-level threshold statements into measurable, interdependent variables. To inform the nodes needed to determine the current state of the risk threshold with respect to social engineering, we review AI's impact on social engineering using a variety of sources. The next section illustrates how a BN could be instantiated for a specific social engineering risk domain.

In practice, each node in the BN corresponds to a variable influencing AI-augmented phishing risk, such as model accessibility, the sophistication of generated content, or the effectiveness of countermeasures. The conditional dependencies between these nodes capture how changes in one factor, such as the model's ability to produce highly convincing messages and leverage external tools or information sources, can alter the probability of others, such as attack success rates. By mapping and quantifying these relationships, the BN provides a structured means to estimate the probability of crossing a defined risk threshold.

# Risk Domain Analysis:
# AI-Augmented Social Engineering

Social engineering attacks exploit vulnerabilities in human psychology by manipulating individuals into sharing information or taking actions that advance a threat-actor's goals. These attacks have entered into a new era with the advancement of AI, which has altered the threat landscape by enhancing the realism, personalization, scope, and scale of attacks, making them harder to detect and defend against.

AI's impact on social engineering is a top concern for those charged with modeling and mitigating the risks posed by frontier systems. Experts have noted that it is imperative to develop a thorough understanding of how AI is impacting the risk landscape today, as well as methods to predict where the greatest impact will be felt in the future (UK AI Safety Institute [UK AISI], 2025). Today, synthetic content produced by generative AI is being used by criminals to perpetuate fraud.[9] Europol, the FBI, and ENSIA have identified several ways criminals are using AI to create realistic synthetic content for a wide range of scams (ENSIA, 2024; FBI, 2024a; Europol, 2025). These include:

- Creating fraudulent images that support criminal activities, including realistic profile pictures for fake social media accounts, fraudulent identification documents like driver's licenses, images of celebrities for counterfeit product promotions or charity scams, and images of everyday victims for sextortion schemes.
- Cloning voices to impersonate people for financial gain. Examples include generating a short audio clip of a loved one to elicit urgent financial assistance or a ransom payment, or using a victim's voice to gain access to their bank accounts.
- Generating convincing video content, particularly for purposes of impersonation and fraud. This includes generating deepfake videos of public figures or authority figures, creating fake videos for real-time video chats to support a fraudulent online identity, and producing misleading promotional videos for investment scams.
- Automating phishing messages to run large-scale scam operations by targeting multiple victims in different languages and using personal information, such as victims' interests,

---

9    NIST defines synthetic content as "information, such as images, videos, audio clips, and text, that has been significantly altered or generated by algorithms, including by AI," and offers guidance on reducing the risks posed by this content in NIST AI 100-4 (NIST, 2024b).

close connections, and home addresses, to create manipulative narratives and tailor
communication with victims to increase the success of romance and investment fraud.

AI developers and deployers themselves have reported extensively on the impact of AI on
social engineering. In August 2023, Google Cloud Mandiant reported that threat actors were
using generative AI, albeit in very limited situations and with limited expertise. They observed
threat actors creating deepfakes and operations in which generative adversarial networks
(GANs), in particular, were leveraged to generate profile photos for inauthentic personas.
Actors then obfuscated the AI-generated origin of their profile photos through tactics such
as adding filters or retouching facial features (Cantos et al., 2023). The company also reported
that financially motivated actors were using manipulated video and voice content in business
email compromise (BEC) scams (Cantos et al., 2023); North Korean cyber espionage actors
were using manipulated images to defeat know-your-customer (KYC) requirements; and other
APT groups were using voice-changing technology in social engineering campaigns targeting
Israeli soldiers (Cantos et al., 2023).[10] The company theorized that, in the years to come, generative
AI would significantly amplify threats in the social engineering domain (Cantos et al., 2023).

This proved to be correct: in January 2025, the Google Threat Intelligence Group (which
includes Mandiant and Threat Analysis Group teams) reported on a very different threat
landscape (Google Threat Intelligence Group, 2025). They stated that, across various APT
groups, threat actors were observed using generative AI for a wide range of tasks, including
vulnerability research, crafting phishing campaigns, conducting reconnaissance on defense
experts and organizations, scripting and development, troubleshooting code, researching
how to obtain deeper access to target networks, lateral movement, privilege escalation, data
exfiltration, detection evasion, converting publicly available malware into another coding
language, and adding encryption functions to existing code, among other tasks (Google Threat
Intelligence Group, 2025).[11] It is notable that in all of the reported use cases, threat actors were
using Gemini alone, unlike the report from 2023, which was a broad overview of the AI misuse
threat landscape. Google is not alone, and other companies such as Anthropic and OpenAI,

---

10      These reported incidents were not related to the misuse of Google products or platforms, but rather emerged as part of a
general threat landscape analysis.

11      Google Threat Intelligence Group also provided an updated description of the unique activities of North Korean APTs in
particular: "North Korean APT actors used Gemini to support several phases of the attack lifecycle, including researching potential
infrastructure and free hosting providers, reconnaissance on target organizations, payload development, and assistance with
malicious scripting and evasion techniques. They also used Gemini to research topics of strategic interest to the North Korean
government, such as the South Korean military and cryptocurrency. Of note, North Korean actors also used Gemini to draft cover
letters and research jobs—activities that would likely support North Korea's efforts to place clandestine IT workers at Western
companies" (Google Threat Intelligence Group, 2025).

among others, have all independently reported on the unique ways threat actors are using their tools to advance their social engineering attacks (Anthropic, 2025d; OpenAI, 2025a).

As a result, we believe social engineering is one of the most important domains to explore rigorous applications of risk thresholds. In order to quantify the elements in this domain for a BN, a thorough analysis of AI's impact on social engineering must be conducted. Schmitt and Flechais (2024) establish a framework to contextualize how AI enhances social engineering through three pillars:

1. **Realistic content generation**, including context-aware text, cloned websites, and synthetic media such as deepfake images, voices, and videos;
2. **Advanced targeting and personalization**, enabled by large-scale data scraping and behavioral analysis; and
3. **Automated attack infrastructure**, where AI generates, distributes, and adapts campaigns at scale, maintains ongoing interactions through automated responses, and employs feedback loops to refine tactics while reducing detection.

Through this lens, we analyze recent developments and AI's impact on social engineering. We combine various pieces of information in order to construct a list of variables, which will later be nodes in a network, and establish the connection between these variables. Additionally, we look for any indications of changes in attack vector methods and provide some thoughts on recent changes in the volume and types of attacks. The phishing case study below demonstrates that this type of decomposition — from thresholds to variables to nodes with assigned probabilities — is possible. It allows us to identify candidate variables and causal pathways that could form nodes and edges within the BN, showing how qualitative threshold elements can be operationalized into a probabilistic model.

# Case Study:
# Phishing Risk Subdomain

Phishing, one of the most common initial access vectors used by external threat actors, can be augmented by AI, increasing the scale and success rates of malicious attacks (Verizon, 2025). Phishing is defined as "the practice of sending e-mails that appear to be from reputable sources with the goal of influencing or gaining personal information" (Hadnagy & Fincher, 2015).[12] A typical phishing attack begins with deceptive communication containing malicious requests, attachments, or links. These scams aim to obtain credentials, sensitive information, personal data, or intellectual property, infect devices, enable financial transactions, or seize control of digital accounts (Hadnagy & Fincher, 2015). All phishing scams share similar strategies for delivery and psychological manipulation as the means to engineer their victims' engagement. Where they differ is the level of targeting, attack vectors, delivery methods, and end goals of the attacks.

AI has the ability to aid at each phase of the phishing attack process, from reconnaissance to delivery (Schmitt & Flechais, 2024). We investigate the effects of AI-enabled phishing on the current threat landscape across several dimensions, including phishing types, threat actors and motivations, attack vectors, and targets.

## CURRENT STATE OF PHISHING

Phishing is a highly effective method of social engineering. It is estimated that in 2023, over 70% of organizations experienced bulk phishing, spear phishing, business email compromise (BEC), smishing, and social media scams (Proofpoint, 2024). Of the successful attacks, over 30% resulted in either the loss of data and/or intellectual property or a ransomware infection (Proofpoint, 2024). The Federal Bureau of Investigation's Internet Crime Complaint Center

---

12      Several subtypes of phishing exist including: (1) email phishing: the most common form of phishing, where deceptive communication is sent via email with the goal of getting a click, such as redirecting the victim to a malicious site or interacting with malicious attachments, or motivating them to take further action to reveal credentials or sensitive information; (2) spear phishing: highly customized and targeted phishing, typically including personal or context specific details; (3) smishing: phishing via SMS or text messaging; (4) whaling: spear phishing targeting significant or high-profile victims (e.g., influential figures, tech CEOs); (5) business email compromise: scam where criminals impersonate trusted figures within an organization to trick other employees; (6) angler / social media phishing: the use of social media to deceive targets; and (7) QR code phishing: the use of malicious QR codes, which are particularly good at evading automated detection and hiding their malicious content (APWG, 2025; Schmitt & Flechais, 2024).

Annual Report for 2024 noted phishing/spoofing as the highest crime type by complaint count, with 193,407 complaints and over $70 million in reported losses (FBI, 2024b). A UK study revealed that in 2024, phishing dominated the threat landscape, with 84% of businesses and 83% of charities reporting phishing attacks (DSIT, 2024a).

We consider the existing prevalence of phishing attacks and the importance of establishing baselines for phishing before and after the release of popular GPAIS/generative AI systems. Grounded in this research, we can begin to look at the unique impact of AI on this attack vector.

## THE EFFECT OF AI ON PHISHING

The success of phishing attacks resulting from generative AI use can be attributed to the three primary pillars discussed above: (1) realistic content generation, (2) advanced targeting and personalization, and (3) automated attack infrastructure (Schmitt & Flechais, 2024). We also look for any indications of changes in the volume of attacks that can be attributed to AI augmentation.

### Realistic Content Generation

AI can be used to craft convincing emails, text messages, or other types of written communication that replicate tone and style and incorporate context-specific details to increase persuasiveness (Schmitt & Flechais, 2024). For example, generative AI is likely a contributing factor in the global increase in BEC attempts, as it enables translations that not only produce perfect spelling and grammar, as traditional translation tools do, but also show an understanding of regional context and tone (Proofpoint, 2024).

Generative AI can also be used for more technical pieces of the kill chain, such as weaponization and delivery. Models have been used to clone websites to create phishing pages that are indistinguishable from trusted websites (Begou et al., 2023) and to generate a variety of both standard and evasive phishing attacks, including those with QR codes, iFrame injection/clickjacking, ReCAPTCHA evasive attacks, polymorphic URL generation, and browser-in-the-browser exploits, among others (Roy et al., 2024). Popular pre-trained models (GPT-3.5, Turbo, GPT-4, Claude, and Bard) used without the need for jailbreaks were found to "convincingly imitate well-known brands and also deploy a range of evasive tactics that are used to elude detection mechanisms employed by anti-phishing systems" (Roy et al., 2024). Although these vulnerabilities were disclosed to developers, the sheer number of possible prompts that could elicit this behavior while bypassing detection, combined with the increased capabilities of open-source models, suggests that it is still likely possible to produce malicious content with popular secured models hosted by developers.

## Advanced Targeting and Personalization

AI scraping can efficiently compile information about potential targets from social media websites, public databases, and other sources, further personalizing communications. AI's ability to analyze large datasets can then be used to predict behavior patterns and identify potential targets. One of the earliest demonstrations of the effectiveness of using AI for this piece of a phishing operation was conducted by Seymour and Tully (2018), who showed that K-means clustering and long short-term memory (LSTM) models could be used to (1) identify high-value targets who were more likely to be susceptible to phishing and (2) generate personalized content from topics on their social media timeline (Seymour & Tully, 2018). In one experiment, the automated tool achieved a click-through rate between 30% and 66%, nearly tripling the success rates of traditional attacks (Seymour & Tully, 2018). A more recent proof of concept aimed at fully automating the reconnaissance and outreach phase of phishing campaigns that utilized open-source intelligence (OSINT) achieved a 54% click-through rate, on par with the human experts that were tested, significantly higher than the 12% success rate for traditional attacks (Heiding et al., 2024a). In prior work modeling AI's theoretical impact on the broader threat landscape, only optimistic predictions of capabilities resulted in  scenarios where AI was on par with human experts at a 30% click-through rate (Lohn & Jackson, 2022).

## Automated Attack Infrastructure

Attackers may also leverage AI capabilities to automate the generation and distribution of phishing attacks or phishing kits, allowing for large-scale attacks with reduced effort. Generative AI systems can be used in tandem, with one generating malicious prompts that are then fed into another model, prompting or jailbreaking it to produce phishing content (Roy et al., 2024). This recursive process significantly reduces the effort required by attackers, enabling them to scale attacks rapidly. In one study, the authors found that it took, on average, 29 seconds to create a basic phishing page, including "cloning the original site, removing original sign-in buttons (e.g., Google Sign-In), cleaning forms, and integrating our Telegram API" (Begou et al., 2023). They were able to generate phishing websites 31.25% of the time, and were unsuccessful in generating more, in part, due to the context window limitations of GPT-3.5-

turbo-16K, which had a context window of 16,384 tokens. The latest GPT, GPT-5, has a context window of 400,000 tokens, which is more than 24 times larger (OpenAI, 2025c).

## Increase in Volume and Scope of Phishing

While the continued appeal of phishing to cybercriminals is predictable, the dramatic escalation in phishing attack frequency and change in targets over the past five years may be cause for concern. According to the Anti-Phishing Working Group (APWG), 1,130,393 phishing attacks were observed in Q2 2025, representing a 311% increase from the 274,681 attacks in Q2 2020 (APWG, 2025; APWG, 2020). Beyond existing incentives to increase the scale and scope of operations, AI also reduces technical barriers, potentially creating new economic incentives to use the technology to assist with attacks (Heiding et al., 2024b).

From this research, we can begin to construct a sense of a BN by encoding these findings into nodes and edges that capture elements such as realistic content generation, targeting and personalization, automation, and volume. The following table presents a non-exhaustive list of elements from our analysis.

### Table 4: Sample Insights to Inform BN Nodes on AI-enabled Cyber Threats

*This table summarizes different types of evidence and metrics that could inform nodes in the BN. Qualitative findings (e.g., insights from red teaming exercises or sociological studies on employee susceptibility) provide structured expert judgment or observed patterns that can be translated into probabilistic estimates for BN nodes. Quantitative measures (e.g., capability evaluation results, defensive response times, or efficacy metrics) provide numeric inputs that directly inform the likelihoods in the BN. Together, these inputs allow the BN to estimate the probability of crossing a defined risk threshold while capturing the uncertainty inherent in both qualitative and quantitative sources.*

| Qualitative Finding | Quantitative Measures |
| --- | --- |
| Realistic Content Generation | *Capability Evaluations*<br>AI evaluations focused on rating models on widely accepted and standardized benchmarks<br><br>*Red Teaming*<br>Red teaming/adversarial tests aimed at assessing the strength of bypass refusal policies and content filters |
| Advanced Targeting and Personalization | *Capability Evaluations*<br>AI evaluations focused on rating models on widely accepted and standardized benchmarks<br><br>*Threat Landscape Assessment*<br>Data from threat intelligence feeds can provide the most up-to-date information on threat actor TTPs |

| Automated Attack Infrastructure | *Capability Evaluations*<br>Specialized evaluations to test a model's multi-step planning, tool use, and autonomous execution |
|---|---|
| | *Threat Landscape Assessment*<br>An evaluation of black market platforms and the availability and cost of malicious AI "as-a-service" tools |
| Increase in Volume and Scope of Phishing | *Historical Data*<br>Prior data focused on capturing the volume of phishing across different types of attacks |
| | *Threat Landscape Assessment*<br>An assessment of threat actor affordances, such as access to powerful models |
| | *Sociological Studies*<br>The uplift given to threat actors at different levels of expertise to quantify how many more threat actors we might expect to move up in capability level |

## AI-Enabled Phishing Countermeasures

While offensive AI poses a major threat, advanced defensive AI also becomes overwhelmingly effective at scale. In the same study that demonstrated that AI models were on par with human experts at crafting successful spear-phishing messages, the authors also showed that Claude 3.5 Sonnet achieved a true positive detection rate of 97.25%, with no false positives (Heiding et al., 2024a).

However, AI-enhanced threats also weaken the effectiveness of traditional countermeasures, including security awareness, training, and education, which will need to adapt. Additionally, traditional machine-learning approaches for detecting phishing attacks (e.g., random forests and support vector machines) have become increasingly ineffective due to their inability to adapt (Kavya & Sumathi, 2024). In response to these challenges, cybersecurity professionals increasingly turn to sophisticated AI-powered countermeasures.

New AI-based defensive strategies using deep learning models have achieved up to 98.9% accuracy but remain vulnerable to adversarial attacks (Sahingoz et al., 2024). GAN-based systems such as PDGAN have reached 97.6% accuracy by integrating URL data, and ensemble learning techniques help boost detection accuracy and reduce overfitting by combining multiple classifiers (Al-Ahmadi, 2022; Kavya & Sumathi, 2024). These approaches, alongside others (e.g., NLP techniques), demonstrate the potential for countermeasures to match the intensity and capabilities of AI-assisted phishing attacks. Yet, many of these countermeasures are resource-exhaustive and lack robustness, emphasizing the need for continuous innovation to prevent attackers from gaining a decisive technological advantage.

A promising approach for AI-aided phishing countermeasures involves integrating early detection methods with LLMs (e.g., as a third-party plug-in), thereby preventing LLMs from generating phishing content. Roy et al. (2024) demonstrated this approach by developing an automated malicious prompt detection tool based on Bidirectional Encoder Representations from Transformers (BERT), a language model developed by Google; this tool achieved average accuracies of 94% for phishing email prompts and 96% for phishing website prompts. Similarly, studies have shown that detection methods leveraging popular LLMs (e.g., ChatGPT, Claude) can identify malicious intent and suggest preventive actions, even in cases where malicious intent is not immediately apparent (Heiding et al., 2024b).

**Table 5: Sample Insights to Inform BN Nodes on AI-enabled Cyber Defenses**

*This table summarizes different types of evidence and metrics that could inform nodes in the BN. Qualitative findings (e.g., insights from red teaming exercises or sociological studies on employee susceptibility) provide structured expert judgment or observed patterns that can be translated into probabilistic estimates for BN nodes. Quantitative measures (e.g., capability evaluation results, defensive response times, or efficacy metrics) provide numeric inputs that directly inform the likelihoods in the BN. Together, these inputs allow the BN to estimate the probability of crossing a defined risk threshold while capturing the uncertainty inherent in both qualitative and quantitative sources.*

| Qualitative Finding | Quantitative Measures |
|---|---|
| Countering Realistic Content Generation | *Red Teaming*<br>Red teaming focused on the effectiveness of filtering mechanics, synthetic content detection, and other AI-specific detection methods |
| Countering Advanced Targeting and Personalization | *Capability Evaluation*<br>Evaluations specifically aimed at capturing the defense system accuracy drop or the rate of traditional defense systems failing when presented with AI-hyper-personalized content vs. standard phishing<br><br>*Sociological Studies*<br>The effectiveness of retraining employees, given the advancements in AI-generated content |
| Countering Automated Attack Infrastructure | *Capability Evaluation*<br>An assessment of defensive metrics such as Mean Time to Contain (MTTC) or Defensive Automation Response Speed in comparison to state-of-the-art AI automation techniques<br><br>*Threat Landscape Assessment*<br>How often organizations update/retrain their defense models to counter faster or more effective attacks |

| Countering Increases in Volume and Scope of Phishing | *Capability Evaluation*<br>LLM guardrail efficacy metrics, such as refusal rates<br><br>*Resource Cost & Implementation Assessment*<br>While defensive measures may exist, a realistic assessment of their cost and probability of proper application is needed |
|---|---|

## THE FUTURE OF AI-AUGMENTED PHISHING

It may be promising for defensive researchers to continue to investigate the use of LLMs in combination with traditional ML approaches. This research must be coupled with the rapid creation and deployment of open-source detection tools that can be easily integrated in critical sectors. The current state of phishing would suggest we have limited time to prepare for the impact of these attacks and little societal resilience to recover from them at scale.

Looking to the future, one of the greatest risks associated with AI-augmented phishing is the combination of targeted spear phishing with the mass approach of traditional phishing campaigns on social media platforms (Jackson, 2022). This risk manifests when AI models are leveraged for reconnaissance, gathering large amounts of detailed information on targets, as well as targeting, deciding which victims will be most susceptible to an attack. The delivery through social media platforms provides an environment where shortened URLs, casual communication, and outreach by unknown individuals are all more common, potentially lowering the guard of victims. The widespread use of LLMs for simple social botnets has already been documented (Yang & Menczer, 2023). AI social media bots capable of delivering targeted attacks at scale should be continuously monitored, and implementing safeguards used in other domains may help mitigate this risk (Murphy et al., 2025). Another high-impact, high-likelihood risk exists with multi-channel attacks.[13] Future phishing campaigns that seamlessly blend email, SMS (smishing), and voice calls to create more immersive and believable fraudulent narratives at scale could be particularly difficult to counteract.

These future threat scenarios point to the need to construct a network that captures various aspects of the social engineering domain beyond point-in-time, narrowly scoped evaluations. It also provides us with insight into which risk pathways might be most interesting to explore. While the average AI cyber evaluation undertaken by an AI developer may naturally focus on the platform and systems under their control, our analysis shows that testing the impact on various platforms, including social media, would be an important step to better understand the risks that general purpose AI poses.

---

13      An employee might receive a text message appearing to be from their bank, followed by an email from their "CEO" referencing the text, and finally a vishing call to "verify" their identity, with each step reinforcing the legitimacy of the others.

## APPLYING THE ANALYSIS TO A MODEL

Using nodes that capture the threshold elements of both offense and defense, we can begin to construct a network with various states and associated probabilities.

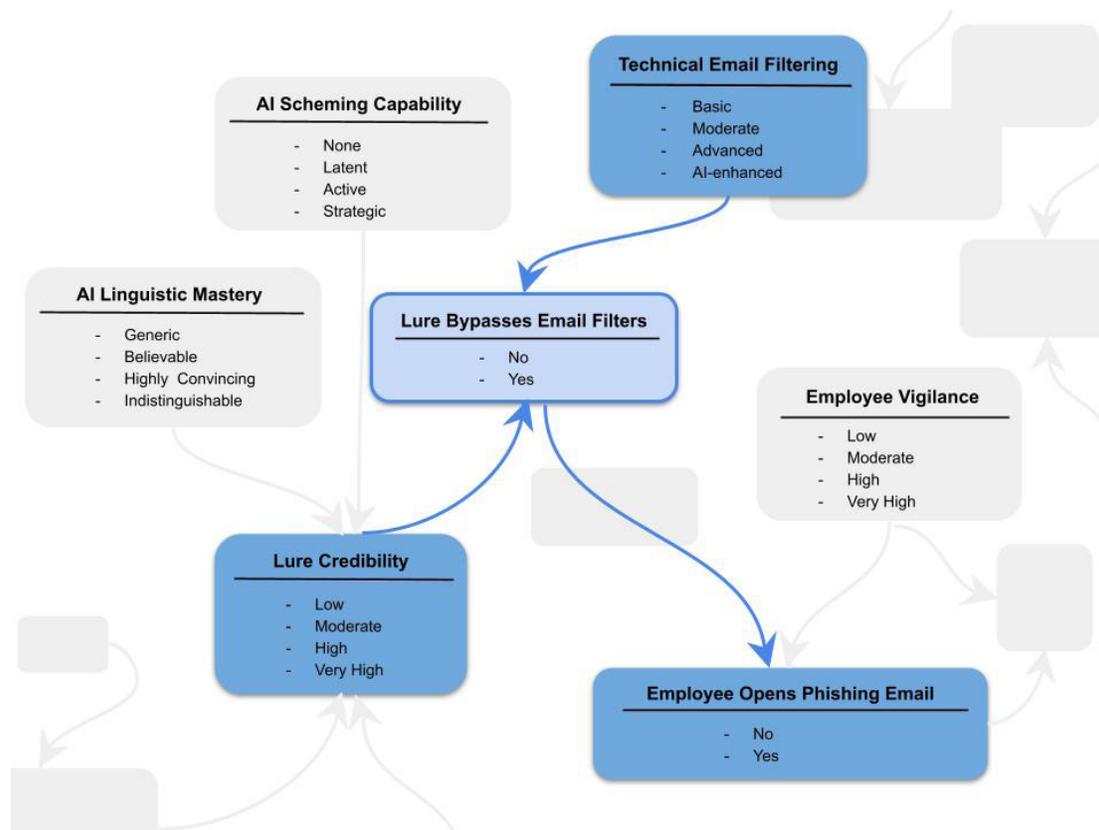

**Figure 3: Illustrative Example of a Bayesian Network for the Phishing Sub-domain**

*This figure offers an illustrative example of what a Bayesian network structure looks like when applied to the phishing sub-domain. The diagram visualizes the conditional dependencies between offensive variables, such as "AI Linguistic Mastery," and defensive mechanisms, like "Technical Email Filtering." By mapping these relationships, the network demonstrates how distinct nodes and their respective states interact to determine the conditional probability of the final outcome, specifically whether an employee opens a malicious email.*

More research is needed to create and test the processes needed to systematically construct a BN for AI-enabled cyber threats. Researchers should first focus on surfacing qualitative findings from diverse sources, including industry reports, user studies, and expert assessments across industry, government, and academic reports, and validating them with subject-matter experts. From there, a descriptive set of nodes and an initial network can be constructed and validated in parallel with efforts to elicit information from experts on the connections between nodes. Informing the nodes is likely to be the most labor-intensive task.

In our investigation into building small-scale, illustrative ("toy") BN models, we found that while a sufficient amount of information exists across industry, government, and academic reports, there is much work to be done to extract and combine the data. Each effort to collect and refine data must account for uncertainty. There may be an important role of experts in validating individual distributions for a given node before they are used as input into a network.

# Conclusion

The evolution from "vibe hacking" to the GTG-1002 campaign demonstrates how AI-enabled threats are advancing. This reinforces the need to move beyond incident response and reporting to proactive governance. Pre-defined thresholds offer one path toward enabling this shift from retrospective analysis to anticipatory oversight.

A central shortcoming of current AI thresholds is that they outline outcomes deemed intolerable but lack robust methods to measure them. Developing credible thresholds will require moving beyond capability-centric indicators toward methods that quantify uncertainty, combine diverse evidence sources, and adapt as capabilities evolve. This paper identifies BNs as one practical route for decomposing high-level threshold concepts, integrating a wide variety of evidence, and updating risk estimates as both AI capabilities and threats evolve.

The case study on AI-augmented phishing demonstrates how heterogeneous evidence can be integrated — specifically, how qualitative threat descriptions can be translated into quantitative variables. This approach allows us to assess how model capabilities, attacker incentives, and defensive measures interact to shape overall risk. But the case study also reveals that despite our ability to identify key variables and map their causal relationships, no standardized method yet exists for translating evaluation results into probabilistic estimates. Our review of methods and examples of BN use across various domains offers an initial starting point for research and further investigation that can be iteratively strengthened. For this reason, the cyber domain, with its mature risk practices and strong data foundations, offers a practical risk area for us to develop methods that can guide threshold development in other high-stakes AI risk areas.

The adoption and validation of BNs will depend on developing elicitation protocols, validation standards, and shared threat models, all of which will be needed to operationalize such tools at scale. This requires investment in high-quality incident data collection, mature evaluation practices, and standardized reference scales for assessing likelihood and impact. Alongside these methods, there is also a need for governance that specifies how probabilistic outputs should trigger concrete actions, such as enhanced monitoring, deployment restrictions, or additional mitigations.

Ultimately, the development of robust risk thresholds will require sustained collaboration across model developers, evaluators, and oversight bodies. Despite their limitations,

BNs represent one possible path forward in the probabilistic modeling of AI risks. When strengthened with sufficient methodological research, they may serve as the most practical and tractable approach for developing defensible and measurable thresholds that can support clearer decision-making under uncertainty and improve our ability to anticipate and mitigate intolerable AI-enabled cyber risks.

# Acknowledgments

This white paper benefited from feedback and informal discussions with colleagues and practitioners. We are grateful to Matthew Broerman, Andrew Lohn, Richard Mallah, James Sykes, Anna Wisakanto, and others for their thoughtful comments and suggestions. The views expressed in this paper are solely those of the author(s), and individuals acknowledged do not necessarily endorse the analyses, recommendations, or conclusions presented here.

# Appendix

Assessing cyber risks posed by advanced AI models requires systematic, repeatable, and contextually relevant evaluation methods that can identify dangerous thresholds before escalation occurs. This appendix catalogues prominent cyber capability benchmarks and evaluation frameworks used to assess the security risks of advanced AI models, with a focus on the types of tasks covered and their relevance for risk assessment across the model lifecycle.

## AI CYBER BENCHMARKS

| # | Name | Type | Description | Range of Tasks Covered / Pros |
|---|------|------|-------------|-------------------------------|
| 1 | CYBERSECEVAL 3 | Benchmark | Designed to empirically score model risk across categories, including offensive AI, automated social engineering, vulnerability ops, and agent autonomy. Covers both third-party and end-user risks. | Wide risk coverage, benchmarks SOTA models, and empirical scoring. Useful for quantifying model capabilities across multiple threat scenarios. |
| 2 | CyberGym | Evaluation Framework | Tests AI agents on over 1,500 real-world vulnerabilities from major software projects. It focuses on generating proof-of-concept exploits from vulnerability descriptions and source code, emphasizing agent autonomy and reasoning. | Extensive coverage of real-world codebases, assessment of exploit generation and autonomous agent reasoning. Valuable for understanding model capabilities in practical scenarios. |
| 3 | NYU CTF Dataset | Dataset | Consists of 200 validated CTF problems across six categories to test hands-on cyber reasoning and exploit skills. | Diverse, practical, validated challenges; effective for evaluating both offensive and defensive skills in a controlled environment. |
| 4 | DebugBench | Benchmark: CTF/QA | Focused on bug-hunting and vulnerability debugging using trace analysis, covering 22 bug categories. | Focused software exploit/trace analysis, practical debugging. |
| 5 | SecQA | QA Benchmark | Benchmark with over 2,000 multiple choice questions on security protocols, malware, and process forensics. | A wide range of security topics helps assess knowledge breadth. |
| 6 | OpsEval | Cyber Range | Multi-step offensive and defensive scenario simulation evaluating agent skill and performance over time. | Complex chain attack and defense evaluations, providing insights into time-based performance and operational nuance. |

| 7 | CYBENCH | Benchmark: CTF/Agent | Featuring 40 CTF tasks for agent-based code exploit and analysis. | Tests agent autonomy and exploit generation in a variety of web and application domains, which is useful for assessing offensive capabilities. |
|---|---|---|---|---|
| 8 | XBOW Validation Benchmarks | Pen-test Suite | Provides 104 web security challenges for autonomous penetration testing of agents and multi-agent frameworks, emphasizing reproducibility and isolation. | Offers a variety of web security scenarios, including SSRF, XSS, and injection attacks, and uses Docker isolation for reproducibility, which enhances reliability. |
| 9 | SECEVAL | QA/Agent/ Multi | Secure systems engineering evaluation that covers process, threat, vulnerability, and mitigation, using a mix of agent and QA methods. It offers over 2000 multiple-choice questions across nine domains. | Provides system context and links threats to vulnerabilities, offering a multi-method approach that can assess both knowledge and operational skills. |
| 10 | CYBERBENCH | Benchmark: CTF/QA | Benchmark for offensive and defensive shellcode, reconnaissance, and attack pivoting, designed to test multi-domain cyber skills. | Covers a range of skill domains and uses validated CTF-style challenges, making it effective for targeted skill assessment. |
| 11 | CYBERMETRIC | Metric Tool | Tool that quantifies cyber risk trends and model performance using probabilistic analysis. | Provides risk scoring and trend tracking; valuable for understanding long-term changes in model performance and risk exposure. |
| 12 | SWE-bench Verified | Debug/Patch | Evaluates secure code patching by verifying fixes in real bug reports, focusing on real-world vulnerabilities. | Offers practical evaluation of patch verification and vulnerability remediation, important for assessing defensive capabilities. |
| 13 | BountyBench | Bug Bounty | Uses bounty-style reward tasks to test adversarial skills in finding vulnerabilities, simulating real-world bug bounty scenarios. | Incentivized discovery approach encourages practical exploit finding and can reveal model strengths in adversarial contexts. |
| 14 | Occult | Multi-Agent Benchmark | Evaluates operational adaptability and emergent strategy in cyber scenarios, focusing on agent interaction. | Allows for assessment of emergent strategies and adaptability in operational contexts; valuable for understanding complex agent behaviors. |
| 15 | CRUXEval | Red/Blue Team | Realism-focused benchmark for multi-domain red/blue team challenges. | Evaluation of multi-chain attacks and defense strategies, providing insights into agent complexity and operational effectiveness. |

| 16 | AutoPenBench | Pen-test | Evaluates autonomous penetration testing abilities on varied targets, including exploit chain capability and persistence. | Tests automation and offensive scale important for understanding model persistence and adaptability in attack scenarios. |
|---|---|---|---|---|
| 17 | RedCode | Attack Sim | Simulates realistic code compromise using LLM-driven adversarial input, covering 25 types of vulnerabilities with over 4,000 test cases. | Provides targeted evaluation of offensive capabilities and code-specific exploit generation; useful for risk quantification. |
| 18 | SeCodePLT | Platform QA | Benchmarks security scenarios across multiple code languages and toolchains, emphasizing platform integration. | Multi-language and platform integration scenarios allow for broad assessment of model capabilities in different environments. |
| 19 | MITRE ATT&CK by GPT-3.5 | Scenario QA | Maps model performance to MITRE ATT&CK TTPs. | Mapping to TTPs provides insights into procedural knowledge and adversary emulation. |
| 20 | LLMs for CTF & Certs | CTF/QA | Evaluates LLMs on CTF and certification questions, testing practical cyber skills in challenge-solving scenarios. | Variety of challenges and practical skill evaluation make it useful for assessing foundational offensive capabilities. |
| 21 | DeepMind Dangerous Capability Evaluations | Framework | Assesses dangerous capability spectrum and escalation risk for SOTA models. | Broad risk spectrum and frontier model review provide valuable context for threshold-setting and risk escalation. |
| 22 | LLMs Killed the Script Kiddie | Scenario/Meta | Meta-analysis of AI agents' uplift in threat testing and cyber attack proficiency. | Analysis of workflow changes and attacker skill enhancement provides insights into the evolving threat landscape. |
| 23 | Offensive Security Evaluations | Evaluation/QA | Directly tests LLMs on offensive security challenges | Breadth of challenge-solving and adversarial reasoning make it effective for evaluating offensive capabilities. |
| 24 | CyberSOCEval | QA Benchmark | Focuses on malware analysis and threat intelligence reasoning to evaluate LLMs using a multiple choice question/answer format. | Evaluates the ability to detect threats accurately and map attack chains to frameworks for SOC analysts, which is useful to address evaluation gaps in cybersecurity operations. |

| 25 | SWE-Lancer | Benchmark | Evaluates models on over 1,400 freelance software engineering jobs, encompassing independent engineering tasks and managerial tasks | Provides a more representative and comprehensive evaluation by using real-world data and E2E tests created by professional engineers and involving tasks spanning diverse categories. |
| --- | --- | --- | --- | --- |
| 26 | CTIBench | Benchmark | Assesses model performance in CTI applications by constructing a knowledge evaluation dataset with multiple-choice questions. | Focuses on evaluating LLMs' comprehension, reasoning, and problem-solving capabilities in a broad domain of CTI, which is useful to understand practical applications of LLMs in CTI. |
| 27 | SECURE | Benchmark | Evaluates LLMs performance in cybersecurity scenarios using six comprehensive datasets. | Offers insights and recommendations on application of LLMs in cybersecurity tasks and how to improve their practical usability. |
| 28 | SecLLMHolmes | Framework | Investigates whether LLMs can reliably identify and reason about security bugs by constructing a set of 228 code scenarios, spanning eight critical vulnerabilities. | Fully automated framework to test LLMs' ability, which helps answer whether they can be used as security assistants for vulnerability detection. |
| 29 | NetSPI | Framework | Analyzes how LLMs handle adversarial conditions and measures impact on usability. | The dual approach balances security with functionality, which provides insights on how to improve model robustness without compromising performance. |

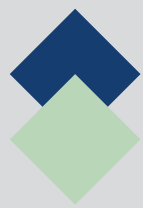

# CLTC

## Center for Long-Term Cybersecurity

UC Berkeley